\documentclass{elsart}

\usepackage[dvips]{epsfig}
\usepackage{amssymb,amsbsy}
\usepackage{amsfonts,amsmath,bm}
\RequirePackage{graphics,epsf}

\begin{document}

\runauthor{Kunert and Harting}
\begin{frontmatter}
\title{Simulation of fluid flow in hydrophobic rough microchannels}

\author[icp]{Christian Kunert and Jens Harting}
\address[icp]{Institut f\"ur Computerphysik, Pfaffenwaldring 27, 70569 Stuttgart, Germany}



\date{\today}


\begin{abstract}

Surface effects become important in microfluidic setups because the
surface to volume ratio becomes large. In such setups the surface
roughness is not any longer small compared to the length scale of the
system and the wetting properties of the wall have an important
influence on the flow.  However, the knowledge about the interplay of
surface roughness and hydrophobic fluid-surface interaction is still
very limited because these properties cannot be decoupled easily in
experiments.
 
We investigate the problem by means of lattice Boltzmann (LB)
simulations of rough microchannels with a tunable fluid-wall
interaction.  We introduce an ``effective no-slip plane'' at an
intermediate position between peaks and valleys of the surface and
observe how the position of the wall may change due to surface
roughness and hydrophobic interactions.

We find that the position of the effective wall, in the case of a
Gaussian distributed roughness depends linearly on the width of the
distribution. Further we are able to show that roughness creates a
non-linear effect on the slip length for hydrophobic boundaries.

\end{abstract}
\end{frontmatter}

\section{Introduction}
In microfluidic systems boundary effects are significantly more
important because of the large surface to volume ratio.  The influence
of the surface topologies and wetting behavior on the fluid flow is an
important issue in the context of microfluidics and cannot be
neglected, since a flat non-interacting surface is always an
idealization.  In this paper we utilize lattice Boltzmann simulations
to investigate the combined influence of roughness and wettability on
the fluid flow.  In the past the influence of roughens was mainly
investigated in the context of (turbulent) boundary layer theory.
Already in the 1930s Nikuradse performed experiments with sand glued
inside a pipe in order to obtain the so called skin friction
coefficient, which is still commonly used in engineering
applications~\cite{schlichting-79}.  Until today, every known theory
describing general rough surfaces requires at least one empirical
parameter to describe the shape of the
surface~\cite{landau-lifschitz-VI}.

This leads to the question which boundary condition has to be applied
at a surface. For more than a hundred years the no-slip boundary
condition was successfully applied in engineering applications.
Nevertheless, Navier~\cite{navier-23} introduced a slip boundary
condition
\[
v(x=0)=\beta \frac{\partial v}{\partial x}
\] 
saying that the fluid velocity $v$ at the boundary $x=0$ is
proportional to the velocity gradient $\frac{\partial v}{\partial
x}$. The constant of proportionality is given by the slip length
$\beta$. $\beta$ depends on many parameters like the wettability, the
surface roughness or fluid properties like the viscosity or molecular
interactions.  Therefore it has to be seen as an empirical length that
contains many to some extent unknown interactions. However, for simple
liquids the measured slip lengths are commonly of the order of up to
some tens of nanometers.

The influence of roughness on the slip length $\beta$ has been
investigated by numerous authors. On the one hand roughness leads to
higher drag forces and thus to no-slip on macroscopic scales.
Richardson showed that even if on a rough surface a full-slip boundary
condition is applied, one can determine a flow speed reduction near the
boundary resulting in a macroscopic no-slip
assumption~\cite{richardson-73}. This was experimentally demonstrated by
McHale and Newton~\cite{mchale-newton-04}. Jansons has shown analytically
that even few perturbations on flat surfaces lead to mesoscopic
slip~\cite{jansons-87} and therefore macroscopically to no-slip.  On the
other hand, roughness can cause pockets to be filled with vapor or gas
nano bubbles leading to apparent
slip~\cite{du-goubaidoulline-johansmann-04,joseph-etal-06}.  Jabbarzadeh
\emph{et al.}~performed molecular dynamics (MD) simulations of Couette
flow between sinusoidal walls and found that slip appears for roughness
amplitudes smaller than the molecular length
scale~\cite{jabbarzadeh-etal-00}. For the creation of superhydrophobic
surfaces which are reducing the drag enormously, it is essential that
there are variations in the surface height~\cite{herminghaus-00}.

Varnik \emph{et al.}~\cite{varnik-dorner-raabe-06} applied the lattice
Boltzmann (LB) method to show that even in small geometries rough
channel surfaces can cause flow to become turbulent.

Recently, Sbragaglia \emph{et al.} applied the LB method to simulate
fluids in the vicinity of microstructured hydrophobic
surfaces~\cite{sbragaglia-06}. In an approach similar to the one proposed
in this paper, they modeled a liquid-vapor transition at the surface
utilizing the Shan-Chen multiphase LB model~\cite{shan-chen-93}. The
authors were able to reproduce the behavior of the capillary pressure as
simulated by Cottin-Bizonne \emph{et al.} using molecular dynamics (MD)
simulations quantitatively~\cite{cottin-bizone-etall-04}. They further
showed that hydrophobic rough surfaces increase the mass flow which
corresponds to an apparent slip effect and that there exists a ``critical
roughness'' at which superhydrophobic effects appear.

A common setup to measure slip is to utilize a modified atomic force
microscope (AFM) to oscillate a colloidal sphere in the vicinity of a
boundary~\cite{bib:vinogradova-95,bonaccurso-03,vinogradova-yakubov-06}.
In these measurements the drag force is measured by the AFM and
compared with the theoretical values. By applying the theory of
O.I. Vinogradova to quantify the force correction needed one can find
the corresponding slip length $\beta$ \cite{bib:vinogradova-95}.  Here
the distance between sphere and wall becomes very small. Therefore a
correct knowledge of the boundaries' properties is very important.
Vinogradova and Yakubov demonstrated that assuming a wrong position of
the surface during measurements can lead to substantial errors in the
determined slip lengths~\cite{vinogradova-yakubov-06}. They showed
that measurements can be interpreted by assuming a modified boundary
position instead of Navier's slip condition, so that the position of a
no-slip wall would be between peaks and valleys of the rough
surface. In this paper we follow this idea. We answer the question at
which distinct position the ``effective boundary'' has to be placed
and study the influence of a wrongly determined wall position
numerically.

In previous papers we presented a model to simulate hydrophobic
surfaces with a Shan-Chen based fluid-surface interaction and
investigated the behavior of the slip length
$\beta$~\cite{harting-kunert-herrmann-06,kunert-harting-06}. We showed
that the slip length $\beta$ is independent of the shear rate, but depends
on the pressure and on the concentration of surfactant added. Recently,
we presented the idea of an effective wall for rough channel
surfaces~\cite{kunert-harting-07}.  We investigated the influence of
different types of roughness on the position of the effective boundary.
In this paper we show how the effective boundary depends on the
distribution of the roughness elements and how roughness and
hydrophobicity interact with each other.

This paper is organized as follows: after the introduction we shortly
present the simulation technique and the simulated system. We give an
overview on the analytical results of Panzer and
Liu~\cite{panzer-liu-einzel-92} which can be compared to our simulations.
Then we present new results on flow among Gaussian distributed roughness
and the correlation between non-wetting and roughness. We close with a
conclusion and summary of our results.

\section{The model}
We use a 3D LB model as presented
in~\cite{bib:jens-harvey-chin-venturoli-coveney:2005,harting-kunert-herrmann-06,kunert-harting-07}
to simulate pressure driven flow between two infinite rough walls that
might be wetting or non-wetting. Previously, we applied the method to
study flows of simple fluids and complex mixtures containing
surfactant in hydrophobic
microchannels~\cite{harting-kunert-herrmann-06,kunert-harting-06}.
Since the method is well described in the literature we only shortly
describe it here.

The lattice Boltzmann equation, 
\begin{equation}
\label{LBeqs}
\begin{array}{cc}
\eta_i({\bf x}+{\bf c}_i, t+1) - \eta_i({\bf x},t) =
\Omega_i, &  i= 0,1,\dots,b\mbox{ ,}
\end{array}
\end{equation}
with the components $i= 0$, 1, $\dots$, $b$, describes the time
evolution of the single-particle distribution $\eta_i({\bf x},t)$,
indicating the amount of quasi particles with velocity ${\bf c}_i$, at
site ${\bf x}$ on a 3D lattice of coordination number $b=19$, at
time-step $t$.

We choose the Bhatnagar-Gross-Krook (BGK) collision operator
\begin{equation}
\label{eq:BKG} 
\Omega_i = -\tau^{-1}(\eta_i({\bf x},t) - \eta_i^{
\, {\rm eq}}({\bf u}({\bf x},t),\eta({\bf x},t))),
\end{equation}
with mean collision time $\tau$ and equilibrium
distribution $\eta_i^{\rm eq}$~\cite{harting-kunert-herrmann-06,succi-01}.
As equilibrium function we choose the expansion
\begin{equation}
\label{Equil}
\!\!\!\!\!\! \eta_i^{\rm eq}\! = 
\! \zeta_i\eta \!
\left[\!1+\!
\frac{{\bf c}_i \cdot {\bf u}}{c_s^2}\! +\!\frac{({\bf c}_i \cdot {\bf
u})^2}{2c_s^4}
\!-\!\frac{u^2}{2c_s^2\!}\!+\!\frac{({\bf c}_i \cdot {\bf u})^3}{6c_s^6}
\!-\!\frac{u^2({\bf c}_i \cdot {\bf u})}{2c_s^4}\right]\! \mbox{.}\!\!\!
\end{equation}
%
We use the mid-grid bounce back boundary condition and choose $\tau=1$
in order to recover the no-slip boundary conditions correctly
\cite{He-Zou-Luo-97}. Interactions between the boundary and the fluid
are introduced as described in~\cite{harting-kunert-herrmann-06},
namely as mean field body force between nearest neighbors as it is
used by Shan and Chen for the interaction between two fluid
species~\cite{shan-chen-93,bib:shan-chen-liq-gas,kunert-harting-06}:
\begin{equation}
\label{eq:force}
{\bf F}^{\rm fluid}({\bf x},t) \equiv -\psi^{\rm fluid}({\bf x},t) g_{\rm fluid,\,wall }
\sum_{\bf x^{\prime}}\psi^{\rm wall}({\bf x^{\prime}},t)({\bf x^{\prime}}-{\bf x})\mbox{ .}
\end{equation}
The interaction constant $g_{\rm fluid,\, wall}$ is set to $0.08$ if
not stated otherwise (see~\cite{harting-kunert-herrmann-06} for a
detailed investigation of the influence of different values for the
interaction constant).  The wall properties are given by the so-called
wall density $\eta_{\rm wall}$. This enters directly into the
effective mass
$\psi^i=1-e^{-\frac{\eta^i}{\eta^0}}$,
with the normalized mass $\eta^0 =1$.  With such a model we can
simulate slip flow over hydrophobic boundaries with a slip length
$\beta$ of up to $5$ in lattice
units~\cite{harting-kunert-herrmann-06}. Higher slip lengths are not
easy to realize since the pressure difference between the bulk and at
the vicinity of the boundary becomes too large leading to
instabilities.  It was shown that this slip length is independent of
the shear rate, but depends on the interaction parameters and on the
pressure.

\begin{figure}[h!]
\centerline{\hspace{1.0cm}%
\epsfig{file=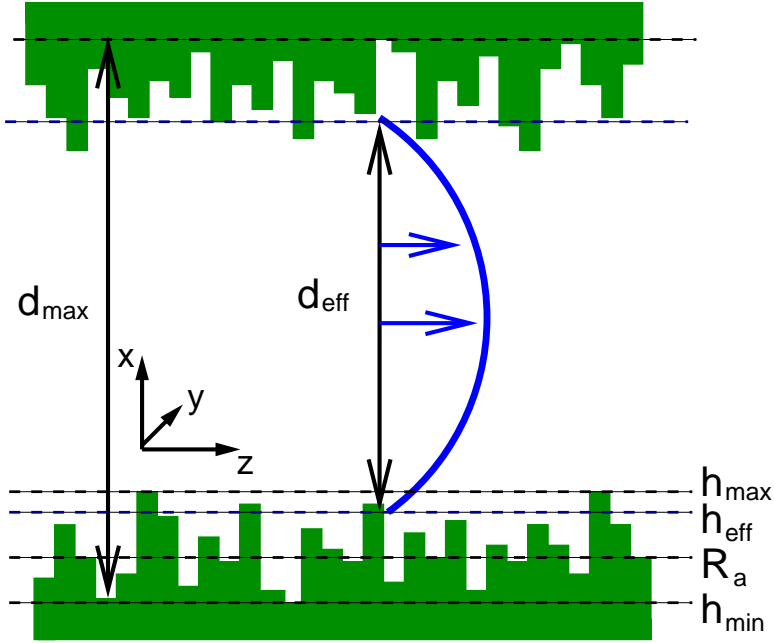,width=0.85\linewidth}}
\caption{\label{fig:hydrodynamicwall} The effective boundary height
$h_{\rm eff}$ is found between the deepest valley at $h_{\rm min}$ and
the highest peak at $h_{\rm max}$ and corresponds to an effective
channel width $d_{\rm eff}$. The baseline is at $h_{min}=0$.  For the
geometries in this paper the average roughness is equal to half the
maximum height $R_a=h_{\rm max}/2$.  The maximum distance between the
plates $d_{\rm max}$ is kept constant of $128$ lattice sites.}
\end{figure}

In this paper we model Poiseuille flow between two infinite rough
boundaries.  Simulation lattices are 512 lattice units long in flow
direction and the planes are separated by 128 sites between the lowest
points of the roughness elements $h_{\rm min}$. Periodic boundary
conditions are imposed in the remaining direction allowing us to keep
the resolution as low as 16 lattice units. A pressure gradient is
obtained by setting the pressure to fixed values at the in- and
outflow boundary.  The highest point of one plane gives the height of
$h_{\rm max}$, while the average roughness is found to be $R_a$ (see
Fig \ref{fig:hydrodynamicwall}). In the case of symmetrical
distributions $R_a=h_{\rm max}/2$.

An effective boundary position can be found by fitting the parabolic
flow profile
\begin{equation}
\label{eq:profil02}
v_z(x)=\frac{1}{2 \mu}\frac{\partial P}{\partial z}
\left[ d^2-x^2-2d\beta \right]
\end{equation}
via the distance $2d=2d_{\rm eff}$. With $\beta$ set to 0 we obtain
the no-slip case. The viscosity $\mu$ and the pressure gradient
$\frac{\partial P}{\partial z}$ are given by the simulation.  To
obtain an average value for $d_{\rm eff}$, a sufficient number of
individual profiles at different positions $z$ are taken into account.
Alternatively, the mass flow $\int v(x) \rho\, {\rm d}x$ can be
computed to obtain $2d_{\rm eff}$.  Both methods are equivalent and
produce identical results. $d_{\rm eff}$ gives the position of the
effective boundary.  The effective height $h_{\rm eff}$ of the rough
surface measured from the minimum height $h_{\rm min}$ is then defined
by $(d_{\rm max}-d_{\rm eff})/2$ (see
Fig.~\ref{fig:hydrodynamicwall}).

\section{Analytic theory}
Panzer \emph{et al.}~calculated the slip length $\beta$ analytically
for Poiseuille flow with rough walls by performing a Fourier expansion
of the streaming function $\Psi$ containing the solution of the
Navier-Stokes equations in the laminar
case~\cite{panzer-liu-einzel-92}.  $\Psi$ is obtained by a Fourier
expansion of the boundary surface and of the pressure field and its
solution contains information of an effective boundary.  The problem
of such an approach is that it works only for small wave numbers.  One
would have to take into account an infinite number of terms to achieve
a result for arbitrary geometries.  Panzer \emph{et al.}~gave an
analytical equation for $\beta$ in the case of small cosine-shaped
surface variations~\cite{panzer-liu-einzel-92}. It is applicable to
two infinite planes separated by a distance $2d$ being much larger
than the highest peaks $h_{\rm max}$. Surface variations are
determined by peaks of height $h_{\rm max}$, valleys at $h_{\rm min}$
and given by $h(z)=h_{\rm max}/2+h_{\rm max}/2\cos(qz)$. Here, $q$ is
the wave number. Since the surfaces are separated by a large distance,
the calculated slip length is equal to the negative effective boundary
$h_{\rm eff}$ that is found to be
\begin{equation}
\label{eq:panzer}
h_{\rm eff}=-\beta=\frac{h_{\rm max}}{2}\left(1+ k\frac{1-\frac{1}{4}k^2+\frac{19}{64}k^4+\mathcal{O}(k^6) }{1+k^2(1-\frac{1}{2}k^2)+\mathcal{O}(k^6)}\right).
\end{equation}
The first and $k$ independent term shows the linear behavior of the
effective height $h_{\rm eff}$ on the average roughness $R_a=h_{\rm
max}/2$.  Higher order terms cannot easily be calculated analytically
and are neglected. Thus, Eq.~\ref{eq:panzer} is valid only for
$k=qh_{\rm max}/2\ll 1$. However, for realistic surfaces, $k$ can
become substantially larger than $1$ causing the theoretical approach
to fail. Here, only numerical simulations can be applied to describe
arbitrary boundaries.

\begin{figure}[h!]
\centerline{
\epsfig{file=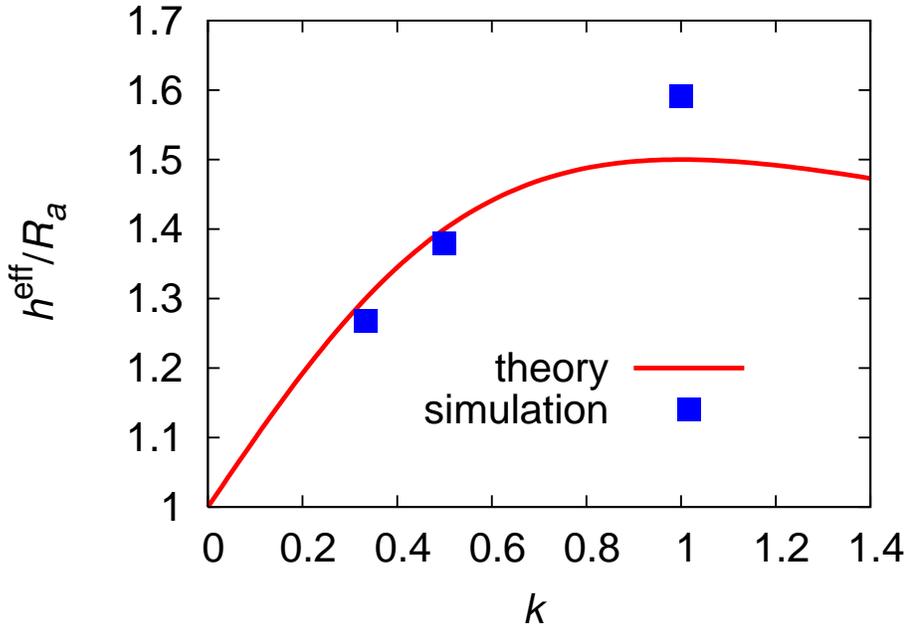,angle=270,width=0.90 \linewidth} }
\caption{ \label{fig:sinus02} Effective height $h_{\rm eff}$
normalized by the average roughness $R_a$ over $k=h_{\rm max}/2 q$ for
a cosine geometry. Symbols denote numerical data and the line is given
by Eq.~\ref{eq:panzer}. For $k>1$ the slope of the theoretical curve
becomes negative, demonstrating that the theory fails for more complex
surface structures while simulations are still valid in this regime.
}
\end{figure}

To test our method we compare our results with the theoretical model
of Panzer and Liu (\ref{eq:panzer}). In Fig.~\ref{fig:sinus02} the
normalized effective height $h_{\rm eff}/R_a$ obtained from our
simulations is plotted versus $k$ for cosine shaped surfaces with
$h_{\rm max}/2=k=1,\frac{1}{2},\frac{1}{3}$ (symbols). The line is
given by the analytical solution of Eq.~\ref{eq:panzer}. For $k<1$ the
simulated data agrees within 2.5\% with Panzer's prediction.  However,
for $k=1$ a substantial deviation between numerical and analytical
solutions can be observed because Eq.~\ref{eq:panzer} is valid for
small $k$ only.  In the case of large $k>1$, the theory is not able to
correctly reproduce the increase of $\beta$ with increasing $h_{\rm
max}$ anymore. Instead, $2\beta/h_{\rm max}$ becomes smaller again due
to missing higher order contributions in Eq.~\ref{eq:panzer}. Our
simulations do not suffer from such limitations allowing us to study
arbitrarily complex surface geometries~\cite{kunert-harting-07}.

\section{Distribution of roughness}
\begin{figure}[h!]
\centerline{
\epsfig{file=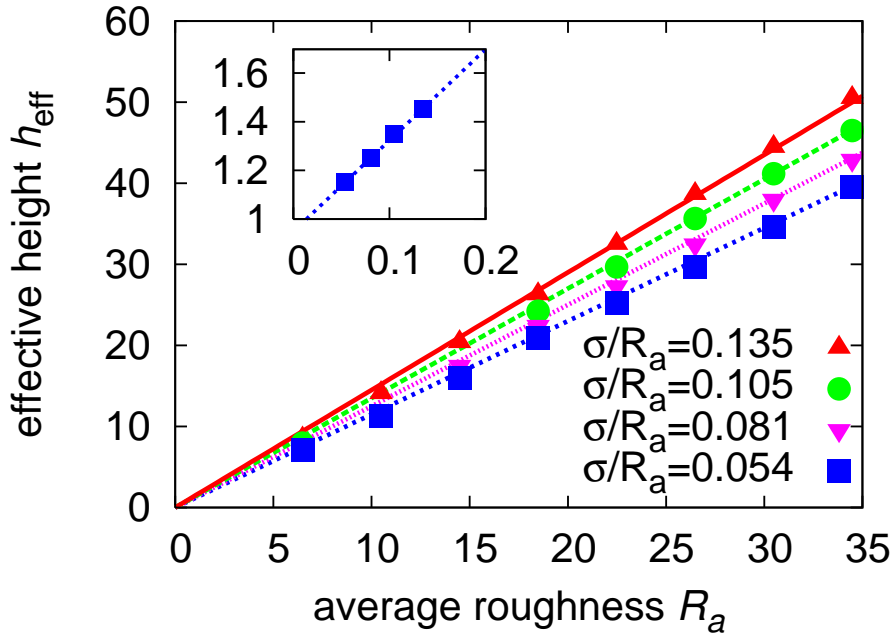,angle=270,width=0.90 \linewidth} }
\caption{ \label{fig:grandom01} Effective height $h_{\rm eff}$ over
average roughness $R_a$ for Gaussian distributed height elements with
different width of the distribution $\sigma$.  Symbols are the
simulation results, lines are a linear fit to the data.  In the inset
the slope of the fitted lines is plotted over $\sigma/R_a$. It shows
the linear dependence of the effective height on $\sigma$.}
\end{figure}

We showed in a recent paper \cite{kunert-harting-07} that the position
of the effective boundary height is depending on the shape of the
roughness elements, i.e., for strong surface distortions it is between
$1.69$ and $1.90$ times the average height of the roughness
$R_a=h_{\rm max}/2$.  By adding an additional distance between
roughness elements, $h_{\rm eff}$ decreases slowly, so that the
maximum height is still the leading parameter. We are also able to
simulate flow over surfaces generated from AFM data of gold coated
glass used in microflow experiments by O.I. Vinogradova and
G.E. Yakubov \cite{vinogradova-yakubov-06}. We find that the height
distribution of such a surface is Gaussian and that a randomly
arranged surface with a similar distribution gives the same result for
the position of the effective boundary although in this case the
heights are not correlated.  To generate this height distribution we
use a Gaussian distributed random height for every lattice point on
the surface obtained from a Box-M\"uller based algorithm
\cite{box-mueller-58}.  We can set the width of the distribution
$\sigma$ and the average height $R_a$. By scaling $\sigma$ with $R_a$
we obtain geometrically similar geometries. This similarity is
important because the effective height $h_{\rm eff}$ scales with the
average roughness in the case of geometrical similarity
\cite{kunert-harting-07}. As an extension of our previous work, we
investigate Gaussian distributed heights with different widths
$\sigma$. In Fig. \ref{fig:grandom01} the effective height $h_{\rm
eff}$ is plotted over the average height $R_a$ for $0.054 <\sigma/R_a
< 0.135$. The height of the effective wall depends linearly on
$\sigma$ in the observed range as it can be seen in the inset.  We
find that the effective height $h_{\rm eff}$ can be fitted by
\begin{equation}
\label{eq:hefflin}
h_{\rm eff}= 1+3.1\sigma .   
\end{equation}
The range of distributions is limited by the resolution of the
lattice. If $\sigma$ becomes too small, the surface is nearly a flat
wall, while a too large $\sigma$ leads to a porous medium instead of a
channel. In order to perform simulations much larger simulation
volumes are needed in the considered case.

The effective height $h_{\rm eff}$ ranges from $1.15 R_a$ to $1.45
R_a$. These values are lower than the effective heights for an equally
distributed roughness ($1.84 R_a$).  We have found in our previous
study that experimentally available surfaces commonly have Gaussian
distributed roughness. Therefore, our results shown here can help to
estimate $h_{\rm eff}$ in real microchannels.

\section{Wettability and roughness}
Next we investigate how roughness and the surface wettability act
together. Therefore we perform simulations with rough channels to
which we assign a fluid-wall interaction as given in the introduction
(\ref{eq:force}).  For $\eta_{\rm wall}$ we choose $0.5$, $1$, and
$5$. For these values and perfectly smooth surfaces we determine the
slip length $\beta$ to be $0.65$, $1.13$ and $1.3$.  As roughness we
choose an equally distributed roughness so that $R_a=h_{\rm max}/2$.
In order to obtain better results we average the values obtained from
four different seeds of the random number generator.  The seed
determines the sequence of random numbers and therefore the shape of
the boundary, by leaving the averaged parameters constant.

\begin{figure}[h!]
\centerline{
\epsfig{file=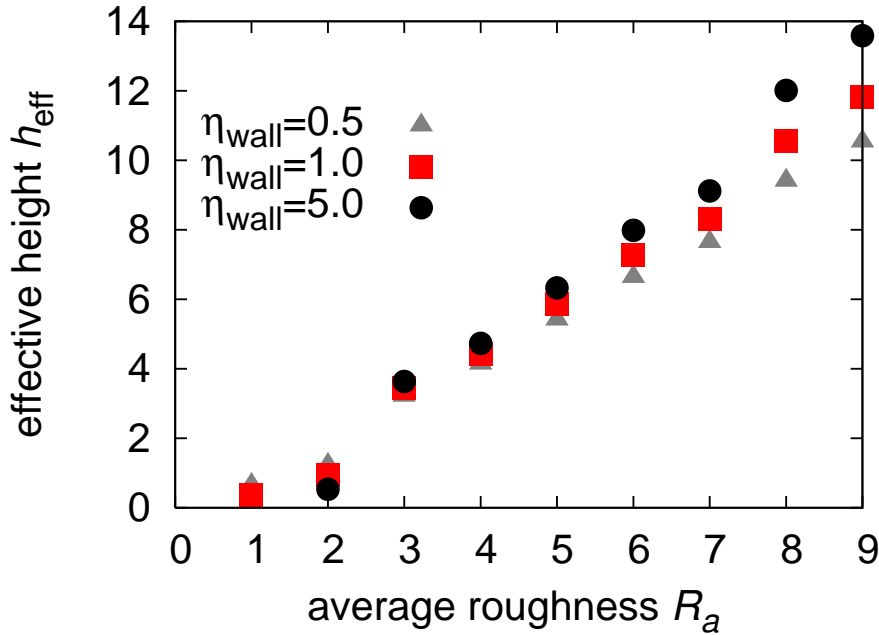,angle=270,width=0.90 \linewidth} }
\caption{ \label{fig:crandom01} Effective height $h_{\rm eff}$ over
average roughness $R_a$ for equally distributed height elements with
different fluid-wall interaction constant $\eta_{\rm wall}=0.5$,
$1.0$, $5.0$. The position of the effective height $h_{\rm eff}$
spreads wider for higher $R_a$, because higher roughness increases the
fluid-wall interaction.  }
\end{figure}

In Fig. \ref{fig:crandom01} we plot the effective height of rough
hydrophobic walls versus the average roughness $R_a$.  For $R_a>4$ we
find a linear dependence between the average roughness $R_a$ and the
effective height $h_{\rm eff}$.  The interesting point is that the
slope for different $\eta_{\rm wall}$ is different. That means that
the fluid-surface interaction dose not cause a simple offset on the
effective height $h_{\rm eff}$ but that a non linear effect is
observed.

\begin{figure}[h!]
\centerline{
\epsfig{file=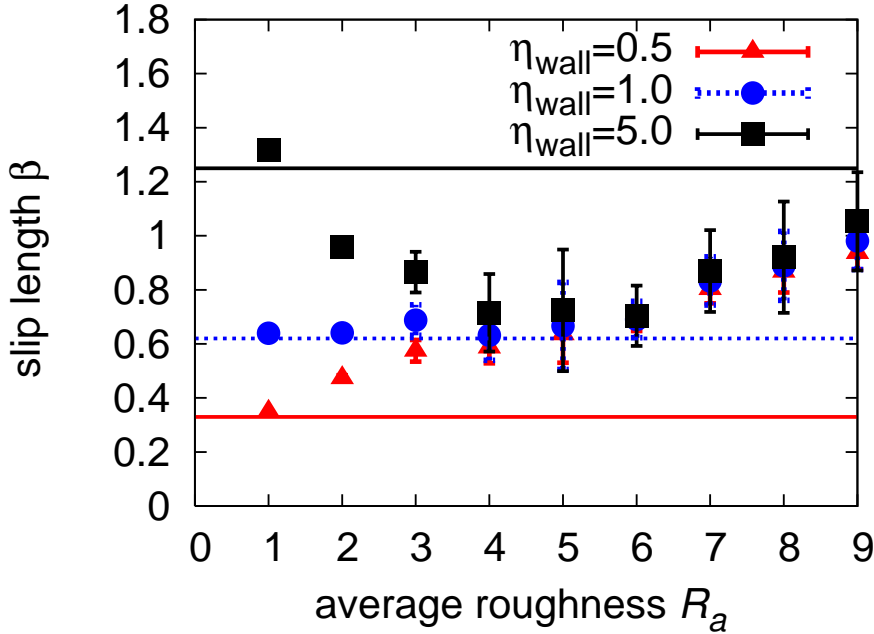,angle=270,width=0.90\linewidth} }
\caption{\label{fig:crandom02} Slip length $\beta$ over average
roughness $R_a$ for equally distributed height elements with different
fluid-wall interaction $\eta_{\rm wall}=0.5$, $1.0$, $5.0$. The
position of the effective height $h_{\rm eff}$ is chosen as the value
for a non-interacting wall. The lines show the slip length for smooth
boundaries ($R_a=0$). Error bars are showing the standard deviation of
results from four different random surfaces.}
\end{figure}

To decouple the effect of roughness and wettability we determine the
slip length by setting the effective distance $d_{\rm eff}$ in
equation (\ref{eq:profil02}) to the effective distance for a rough
no-slip wall.  We then fit the corresponding velocity profile via
$\beta$. In Fig.~\ref{fig:crandom02} we can see that the slip length
$\beta$ for the strong fluid-wall interaction ($\eta^{\rm wall}=5$)
first decreases with the average roughness and then raises.  For a
lower interaction, the slip length is constantly growing and leads to
an increase of the slip length for weak fluid wall interaction
($\eta_{\rm wall}=0.5$) by a factor of more than three.

There are two contradicting effects in this system and their interplay
can explain the observed behavior. The decrease of the slip length
$\beta$ is due to an increased friction near the boundary at moderate
roughness. The increase has its reason in the reduced pressure near
the hydrophobic rough surface, so that the fluid ``feels'' a smoothed
effective surface.  Unlike the implementation of Sbragaglia \emph{et
al.}~\cite{sbragaglia-etal-06} our model is not able to model the
liquid-gas transition near the surface. We only find a density
difference between the bulk fluid and the lattice site just next to
the surface. Therefore, we cannot easily calculate the contact angle
in our simulations~\cite{benzi-etal-06}. However, the general effect
of a pressure drop near a hydrophobic surface is the same as in the
model used in~\cite{sbragaglia-etal-06}. For a more detailed study on
superhydrophobic surfaces, the strong surface variation as well as the
liquid-gas transitions have to be taken into account. This is planned
as a future project and beyond the scope of the current
paper. Nevertheless, it is important to demonstrate the effect of the
pure fluid-surface interaction as it is discussed here.

\section{Conclusion}
In conclusion we performed LB simulations of pressure driven flow
between two rough plates with and without hydrophobic fluid-wall
interaction.  We could show that the effective height scales linearly
with the width of the distribution $\sigma$, so that the effective
height is $h_{\rm eff}=1+3.1 \sigma$.

For rough surfaces with a hydrophobic fluid-wall interaction, we could
show that there exists a strong non-linear effect that leads to an
increase of the slip length $\beta$ by a factor of three for small
interactions.  The behavior of the slip length can only be explained
as a coupled effect since the pure roughness and the pure hydrophobic
interaction were investigated in previous studies.  For further
investigations of super hydrophobicity a phase transition model would
be needed.

We thank H.~Gong for the AFM data and O.I.~Vinogradova and M.~Rauscher
for fruitful discussions. This work was financed within the DFG
priority program ``nano- and microfluidics'' and by the
``Landesstiftung Baden-W\"urttemberg''. Computations were performed at
the Neumann Institute for Computing, J\"{u}lich and at the Scientific
Supercomputing Center Karlsruhe.


\end{document}